\documentclass[10pt,twoside]{article}

\usepackage[english] {babel}

\usepackage {graphicx}
\usepackage {graphics}
\usepackage {floatflt}
\usepackage {latexsym}
\usepackage {caption3} %Пакет для MikTex 2.6, где содержатся команда для замены двоеточия на точку в подписи к рис., например, Рис.1: -> Рис.1.
\usepackage[labelsep=period]{caption}[2007/11/04 v.3.1e]
\usepackage {amssymb}

\def\noi{\noindent}

\renewcommand{\thesubsubsection}%
        {\arabic{section}.\arabic{subsection}.\arabic{subsubsection}.}

\newcommand{\heads}[2]{\markboth{\protect\small\it #1}{\protect\small\it #2}}

\newcommand{\Arthead}[5]{ \setcounter{page}{#4}\thispagestyle{empty}\noi
    \unitlength=1pt \begin{picture}(500,40)

        \put(0,58){\shortstack[l]{\small\it Gravitation \& Cosmology,
                        \small\rm Vol. #1 (#2), #3, pp. #4--#5    \\%No. #3, pp. #4--#5    \\
        \footnotesize {Proceedings of the 5th Inernational Conference on Gravitation and Astrophysics of Asian-Pacific Countries (ICGA-2001), }    \\
\footnotesize {Moscow, 1-7 October 2001}   \\
\footnotesize\copyright \ #2 \ Russian Gravitational Society} }

    \end{picture}
	 }     		%%% #1: volume; #2: year; #3: issue; #4-#5: pages
\def\prepno#1#2
    {\thispagestyle{empty}
    \noi    \unitlength=1mm
    	\begin{picture}(178,10)
            \put(177,15){\llap{\bf #1}}
            \put(177,10){\llap{\small\rm #2}}
        \end{picture}
          }             %%% #1: prepr.no.; #2: journal-ref

\newcommand{\Title}[1]{\noi {\uppercase{\Large #1}}     }%\\}
\newcommand{\Author}[2]{\noi{\large\bf #1}\\[2ex]\noindent{\it #2}   }%\\}

\newcommand{\Abstract}[1]{\vskip 2mm \begin{center}
        \parbox{16.4cm}{\small\noi #1} \end{center}\medskip}

\newcommand{\foom}[1]{\protect\footnotemark[#1]}

\newcommand{\email}[2]{\footnotetext[#1]{e-mail: #2}
		\addtocounter{footnote}{1}}

\heads{A.M.Baranov and D.A.Baranov}
      {Open Universe Model: Description by Mathieu Functions}

\begin{document}
\twocolumn 
[
\Arthead{8}{2002}{Supplement II}{12}{13}

\Title{OPEN UNIVERSE MODEL: DESCRIPTION BY MATHIEU FUNCTIONS \foom 1}%\\

\vspace{.5cm}
   \Author{A.M.Baranov\foom 2 and D.A.Baranov\foom 3}   %% 
{\it Dep. of Theoretical Physics, Krasnoyarsk State University,
79 Svobodny Av., Krasnoyarsk, 660041, Russia}

%\vspace{.3cm}
%{\it Received 11 October 2005}
%
\vspace{.5cm}
\Abstract
{A model of the open Universe described by a conformally flat 4-metric is considered. The gravitational equations with a perfect Pascal fluid as a source are reduced to the nonlinear equation of oscillations. It is proposed to consider this equation as the Mathieu equation encompassing some cosmological models. It is shown that the behaviour of the Universe state function is in qualitative accordance with the Big Bang scenario.}
\vspace{0.5cm}
]

\footnotetext[1]{Talk presented at the 5th Int. Conf. on Grav. and Astrophys. of Acian-Pacific Countries (ICGA-5), Moscow, 2001.} 
\email 2 {alex\_m\_bar@mail.ru}
\email 3 {dmitri\_@mail.ru}

The interest in cosmological models does not fall in the recent decades, in particular, due to multiple observational data, which in accordance with perfecting the instruments and methods of observation are constantly improved. According to the accepted modern views, the matter filling our Universe changed its state during the evolution: from a stage of physical vacuum to an ultrarelativistic stage and further to the Friedmann stage. It means that topical in the theory is a search for exact solutions of the gravitational equations with the matter equation of state depending on a space-time point.

Let us consider a model of the open Universe of Friedmann type, which can be described by the conformally flat metric

$$
\hspace {-35mm}ds^2= e^{2\sigma(S)}\cdot \delta_{\mu \nu}dx^{\mu}dx^{\nu} =
$$
$$
e^{2\sigma(S)}(dt^2-dx^2-dy^2-dz^2) 
\eqno{(1)}
$$ 
with $\delta_{\mu \nu} =diag(1,-1,-1,-1); S^2 = t^2-r^2$ is a square of distance in the Minkowski world,
$r^2 = x^2+y^2+z^2.$

The Einstein gravitational equations are written as 
$$
\hspace {-55mm} G_{\alpha\beta} = -\varkappa T_{\alpha\beta}, 
\eqno{(2)}
$$
with the Einstein tensor $\;G_{\alpha\beta}$ and the energy-momen-\linebreak
tum tensor (EMT) of the Pascal perfect fluid

$$
\hspace {-40mm} T_{\alpha \beta} = \varepsilon \cdot u_{\alpha}u_{\beta}+
 p\cdot b_{\alpha \beta}, 
\eqno{(3)}
$$
where $\varepsilon$ is the energy density,  $p$ is the pressure, 
$b_{\alpha \beta} = (u_{\alpha}u_{\beta} - g_{\alpha \beta})$ is a projector onto the 3-space, 
$g_{\alpha \beta}$ is the metric 4-tensor, $u_{\alpha}= exp(\sigma)\cdot S_{,\alpha}$  is the 4-velosity, Greek indices take the values $\;0,1,2,3.\;$  The velocity of light and the Newtonian gravitational constant are chosen  to be equal to unity.

Substituting the metric (1) into Eqs.(2), we obtain two differential nonlinear equations (for the energy density and pressure) 

$$
\hspace {-37mm}\displaystyle\frac{dy}{dx} \cdot (x \displaystyle\frac{dy}{dx}-y)=
\displaystyle\frac{y^{6}}{12 x^3} \varkappa\varepsilon  ;
\eqno{(4.1)}
$$
$$
\hspace {-52mm}\displaystyle\frac{d^2y}{dx^2}  =
-\displaystyle\frac{y^{5}}{4 x^4} \varkappa p , 
\eqno{(4.2)}
$$
where the new function $y=e^{\sigma/2}\;$ and the new variable $\;x=1/S $ have been introduced.

The last equation can be reduced to a representation of Newton's second law

$$
\hspace {-64mm}\displaystyle\frac{d^2y}{dx^2}  =  F 
\eqno{(5)}
$$
for the potential force  $\;F=-dU/dy\;$ and the potential function $\;U=\Omega^2(x) y^{2}/2,\;$  if we define the right-hand side of Fq.(4.2) as the force $F$ [1].

Thus we have the differential equation of nonlinear oscillations

$$
\hspace {-48mm}\displaystyle\frac{d^2y}{dx^2}+\Omega^{2}(x)\cdot y=0, 
\eqno{(6)}
$$
where the function $\Omega^{2}(x)$ is the squared frequency.

To solve the problem of finding a solution to this equation it is suggested to reduce this equation to the Mathieu equation, extending some special cases obtained earlier,

$$
\hspace {-30mm}\displaystyle\frac{d^2y}{dx^2} + B^2[1+h cos(\gamma x)] \cdot y =0,
\eqno{(7)}
$$
where $B,\, h,\, \gamma $ are constants.

If $\;\Omega^2=0,\;$ i.e. the pressure is absent ($p=0$), we obtain the Friedmann solution for the open Universe filled with incoherent dust, 

$$
\hspace {-58mm} y_F = 1 - Ax, 
\eqno{(8)}
$$
where $A= const$ is determined the by modern value of dust density in the Universe

$$
\hspace {-56mm}\varkappa\varepsilon_{dust} \approx 12 \displaystyle\frac{A}{S^3}.
 \eqno{(9)}
$$

When $\Omega = B^2 =const$, we have an axact cosmological solution [2] for the open Universe filled with dust and electromagnetic equilibrium radiation, 

$$
\hspace {-22mm}y_{BS} = \sqrt{1+(A/B)^2} \cos(Bx -\alpha_0), 
\eqno{(10)}
$$
where $\;tg{\alpha_0} = A/B $, the constant value $\,B\,$ is connected with the density of equilibrium cosmological electromagnetic radiation

$$
\hspace {-50mm}\varkappa\varepsilon_{el-mag} \approx 12 \displaystyle\frac{B^2}{S^4},
\eqno{(11)}
$$
and the solution (10) at large times ($S \rightarrow \infty;\;$ $x \rightarrow 0$) passes through the Friedmann solution (8).

The Mathieu equation (7) may be rewritten in a dimensionless  form by introducing the variable 
$\;\zeta\ = B/S = Bx :$

$$
\hspace {-24mm}d^2y/d\zeta^2+ [1+h cos(\gamma \zeta/B)] \cdot y =0. 
\eqno{(12)}
$$

For solutions of the Mathieu equation ther are resonance regions in accordance with the values of the parameter 
 $\gamma = 2B/n,\; n = 1,2,...\;$ Here we shall choose the first resonance region ($\gamma = 2B$). Then Eq.(12) will be rewritten as 

$$
\hspace {-28mm}d^2y/d\zeta^2+ [1+h cos(2\zeta)] \cdot y =0. 
\eqno{(13)}
$$

In this case, according to Floquet's theorem the solution of the Mathieu equation may be substituted as 

$$
\hspace {-22mm}y=e^{\mu \zeta} ce_1(\zeta,h) - C\cdot e^{-\mu \zeta} se_1(\zeta,h),
\eqno{(14)}
$$
where $\mu$ is a chracteristic parameter, $ce_1$ and $se_1$ are Mathieu's functions having the form 

$$
\hspace {-34mm}ce_1 (\zeta) = \sum_{k}A_{k} cos(2k+1)\zeta ;
\eqno{(15.1)}
$$

$$
\hspace {-34mm}se_1 (\zeta) = \sum_{k}A_{k} sin(2k+1)\zeta .
\eqno{(15.2)}
$$

The Mathieu functions, the signs and constant $C = \mu + A/B$ are chosen here so that in the transition to the modern epoch ($S \rightarrow \infty;\;$ $x \rightarrow 0$) the solution (14) will pass through the Friedmann solution (8), and at $\mu = 0$ and $k =1$ it will be the same as the solution (10). 

By means of parameter selection it is possible to construct the physical behaviour of the function of state 

$$
\hspace {-20mm}\beta(x)=\displaystyle\frac{p}{\varepsilon} =
-\displaystyle\frac{1}{3}
\displaystyle\frac{ x\cdot y \cdot {d^2y/dx^2} }
{ (dy/dx)\cdot (x dy/dx-y) },
\eqno{(16)}
$$
which is at each instant $S$ an equation of state of the Universe.

It is clear that the first approximation of the solution (14) will be written as 

$$
\hspace {-34mm}y \approx e^{\mu \zeta} \cos{\zeta} - C\cdot e^{-\mu \zeta} \sin{\zeta}
$$
$$
\hspace {6mm}= e^{\mu B/S} \cos(B/S) - C\cdot e^{-\mu B/S} \sin(B/S).
\eqno{(17)}
$$

The numerical value of the parameter $\mu = 0.56$ is found for Eq.(13) using a computational method of a characteristic exponent from Hill's determinant [3]. Other parameters may be selected so that we had the qualitative picture of model behaviour of the Universe's function of state according to the Big Bang scenario. At the first instants (when we have a cosmological singularity) the Universe, being initially in the state of physical vacuum ($S =0, \beta \approx -1$), in an inflationary phase is warmed up and reaches the radiation phase ($\beta \approx +1/3$) and then, being expanded and cooled, slowly evolves to the modern epoch 
($\beta \approx 0$).

\end{document}